\documentclass[conf]{new-aiaa}
\usepackage[utf8]{inputenc}

\usepackage{graphicx}
\usepackage{amsmath}
\usepackage[version=4]{mhchem}
\usepackage{siunitx}
\usepackage{longtable,tabularx}
\title{Fast Neural Network Predictions from Constrained Aerodynamics Datasets}

\author{Cristina White\footnote{Research Assistant, Department of Mechanical Engineering, William F. Durand Building, Room 257, 496 Lomita Mall, Stanford University, Stanford, CA 94305.}}
\affil{Stanford University, Stanford, California, 94305}
\author{Daniela Ushizima\footnote{Staff Scientist, Computational Research Division, Lawrence Berkeley National Lab, 1 Cyclotron Rd, Mail Stop 59, Berkeley, CA 94720.}}
\affil{Lawrence Berkeley National Laboratory, Berkeley, California, 94720}
\author{Charbel Farhat\footnote{Chair, Department of Aeronautics and Astronautics, William F. Durand Building, Room 257, 496 Lomita Mall, Stanford University, Stanford, CA 94305, AIAA Member.}}
\affil{Stanford University, Stanford, California, 94305}

\begin{document}

\maketitle

\begin{abstract}
Incorporating computational fluid dynamics in the design process of jets, spacecraft, or gas turbine engines is often challenged by the required computational resources and simulation time, which depend on the chosen physics-based computational models and grid resolutions. An ongoing problem in the field is how to simulate these systems faster but with sufficient accuracy. While many approaches involve simplified models of the underlying physics, others are model-free and make predictions based only on existing simulation data. We present a novel model-free approach in which we reformulate the simulation problem to effectively increase the size of constrained pre-computed datasets and introduce a novel neural network architecture (called a cluster network) with an inductive bias well-suited to highly nonlinear computational fluid dynamics solutions. Compared to the state-of-the-art in model-based approximations, we show that our approach is nearly as accurate, an order of magnitude faster, and easier to apply. Furthermore, we show that our method outperforms other model-free approaches.
\end{abstract}

\section{Introduction}

In machine learning, models are data-driven, whereas in computational fluid dynamics (CFD), models are physics-based and well-defined by a set of partial differential equations. However, the solution---the variation in space and time of fluid state variables such as density, velocity, momentum, and energy---is unknown until a simulation of the model has converged. Automatically iterating through new designs for optimization is often infeasible and datasets of previous simulations are highly constrained in terms of the number of samples because a single simulation can take days or weeks for a single design. We explore whether our model-free method can enable engineers to make use of these constrained datasets to generate accurate approximations to CFD solutions in seconds, given the information they already have from the simulations they have already run. 

\begin{figure}[hbt!]
\begin{center}
\centerline{\includegraphics[width=0.5\columnwidth]{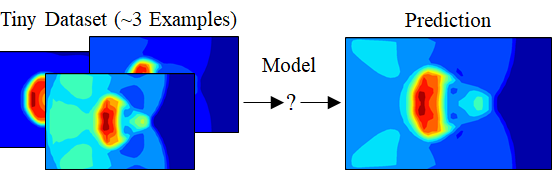}}
\caption{The goal is to automatically build data-driven models that use limited experience to make predictions.}
\label{fig-big-picture-question}
\end{center}
\vskip -0.2in
\end{figure}

Reduced-order models (ROMs) are currently the state-of-the-art model-based methods that address this problem; however, they tend to be highly intrusive. The problem has also been addressed in industry using approximations for scalar 
quantities of interest based on surface fitting techniques: these are fast, but can be inaccurate. Such approaches have also been investigated in combination with neural networks with the aim of improving speed or accuracy, but usually have the same drawbacks as the original approach with which they were hybridized.

Here we address the problem by training stand-alone neural networks to learn the functions that approximate complete solutions instead of quantities of interest. For two different CFD test cases, we reformulate the problems to make 
more efficient use of the data, and we show that fully connected feed-forward networks perform reasonably well, with some room for improvement. We introduce a new neural network architecture, the cluster network, with paired  function and context networks, which addresses the limitations of the fully connected network. The cluster network has a stronger inductive bias--- that the solutions we are approximating are made up of a small number of local, simple functions. For a one-dimensional but shock-dominated problem, we show that the cluster network delivers an accuracy comparable to that of a ROM approach based on a global reduced-order basis, but runs orders of magnitude faster online.

\section{Background and Related Work}

Fluid dynamics introduces a variety of its own unique datasets and problems where machine learning can potentially apply. In the following sections, we map out the areas of intersection between CFD and machine learning and we delineate where our novel proposed method fits.

\subsection{Model-Based Methods}

\subsubsection{Full Order Models}

\paragraph{CFD Solvers:} Industry CFD solvers simulate the Navier-Stokes equations---a set of partial differential equations governing the motion of fluids (both liquids and gases). High fidelity CFD solvers allow for deeper analysis and better aerodynamic design performance. Methods for improving the approximations and performance of the physical models have been investigated for decades with the goal of accelerating simulations. However, due to the high dimensionality required to resolve small scale physics, even the most highly optimized solver may take too long on a supercomputer, which hinders investigations in design and engineering applications. Therefore, machine learning methods for dramatically reducing the computational cost of obtaining solutions while preserving sufficient accuracy are an active and high impact area of research. 

\paragraph{NN Solvers:} Neural network (NN) architectures can be constructed to perform steps that are similar to a CFD solver after training on simulation data. The effort to use neural networks to solve partial differential equations (PDEs) with a focus on fluid dynamics has grown over time \cite{lagaris1998artificial,parisi2003solving,tsoulos2009solving,mall2013comparison}. Eventually, these types of solvers matured into physics informed surrogate deep learning models \cite{raissi2018hidden,raissi2019physics}. These iterative neural network solvers do not gain much in terms of speed, with simulation times on the same order of magnitude as the original solver. But they are a proof of concept that neural networks can approximate CFD solver computations, sacrificing very little in terms of accuracy. These methods struggle to compete with the more than 50 years of optimization behind standard PDE solvers. Instead, methods in PDE solving may be more useful $to$ the machine learning community for faster neural network training \cite{chen2018neural}.

\paragraph{CFD Solver / NN Solver Hybrids:} In CFD and NN hybrid technology, the neural network learns terms that are hard for the CFD solver or unknown, like constitutive material models \cite{javadi2003neural} or turbulence closure terms \cite{ling2016reynolds}. These model additions can improve the accuracy of simulation, while incuring some computational cost. 

\paragraph{CNN Solvers:} Convolutional neural network (CNN) based surrogate models are a clearer win for neural networks---methods in \cite{guo2016convolutional,stoecklein2017deep} were found to dramatically improve solver CPU time, while sacrificing very little accuracy. The CNN surrogate models tend to be a couple orders of magnitude faster than a standard solver. However, these methods can only be applied to spatially uniform grids---a very small subset of aerospace industry problems. One field where these CNN based methods are successfully applied is for the acceleration of computer graphics simulations for Eulerian fluids \cite{tompson2016accelerating}. 

\subsubsection{Reduced Order Models}

\paragraph{ROMs:} ROM and machine learning approaches use similar techniques to create a strategy for reducing the computational cost of CFD while retaining high 
accuracy \cite{amsallem2008interpolation}. ROMs simulate the original governing equations after projecting them onto a low-dimensional solution manifold using a reduced-order basis, and by introducing other approximations into the governing equations such as hyperreduction \cite{uHYP1,GNAT}, reducing computation cost often by several orders of magnitude \cite{uHYP2,MS}. ROMs use knowledge gained from previous simulations to construct the reduced-order basis, and can begin to approximate solutions well starting from a handful of examples---hence, they are often useful for industry problems where getting extremely large CFD datasets for training is infeasible. \cite{carlberg2011efficient,zahr2018efficient}. For training, ROMs utilize an adaptive sampling method known as a ``greedy'' procedure for sampling the parameter space and generating high-dimensional training solution snapshots, which allows them to minimize the training time as much as possible. For highly nonlinear problems, they have been equipped with the concept of local reduced-order bases \cite{LROM}, which has allowed them to accelerate the solution of three-dimensional, turbulent, viscous CFD problems associated with complete aircraft configurations by orders of magnitude, while delivering unprecedented accuracy, including at shock locations \cite{KW,GNAT}. However, they can be intrusive for compressible flows, for example, which can be a significant drawback. 

\paragraph{ROM / NN Hybrids:} In hybrid models, combining ROMs and neural networks, the neural network is trained to model the terms that are hard for the ROM or unknown, resulting in more accurate ROMs with some computational overhead. Some methods may learn the projection errors \cite{san2018neural} or employ neural networks to accurately approximate the coefficients of the reduced model \cite{hesthaven2018non}.

\subsection{Model-Free Methods} 

\subsubsection{Surface Fitting}

Surface fitting on a quantity of interest (QoI) simply maps design parameters directly to QoIs using linear or nonlinear fitting. These methods are simple and not intrusive. Their deterministic versions can be inaccurate \cite{chilenski2015improved}
as they suffer from uncertainty with respect to the choice of the reconstruction data. However, their nondeterministic variants,
known as krieging or Gaussian processes, successfully address this issue. Given large datasets, neural networks have been investigated as alternative methods for mapping, and where large enough datasets are available, have been shown improve upon deterministic curve fitting methods \cite{trizila2011low}. Along with running as many full order models as computationally feasible, various forms of surface fitting constitute the most common methods employed for design optimization in industry. They are employed whenever the analysis of a system can be limited to a static analysis and a few QoIs.

\subsubsection{Solution Fitting}

Solution fitting is our novel proposed approach using neural networks to learn to approximate entire solutions from few examples. The method uses a simple network to learn the function that approximates the CFD solutions in a way that generalizes well from only a few examples. Using this method, solutions for very large number of design parameters can be computed nearly instantaneously with one forward pass of a small network, thereby competing with ROMs. This method is unlike surface fitting in that the network does not only learn a QoI for a given design parameter, but the entire solution, from which any QoI can be computed.

\section{Method}

The proposed architectures for full order solution fitting manage the limitation that few examples of design parameters are available by learning the function that approximates the training set solutions. Variables are mapped from ($x, t, \mu$) values, which are known, to a fluid state variable of interest, $u$, which is unknown. Once either network architecture is constructed, it is trained on the dataset. For evaluation, predictions for unseen design parameters in the test set are computed, and percent error is tabulated and compared to baselines. 

\subsection{Cluster Network} 

\begin{figure}[hbt!]
\begin{center}
\centerline{\includegraphics[width=0.43\columnwidth]{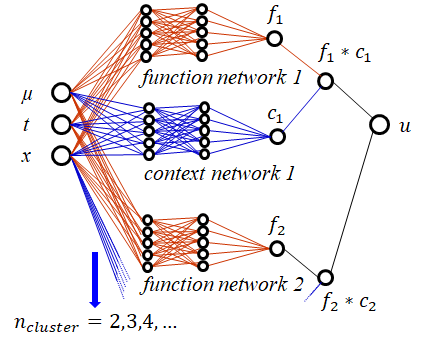}}
\caption{Cluster network.}
\label{fig-clusternet}
\end{center}
\vskip -0.2in
\end{figure}

The novel proposed architecture is shown in Figure~\ref{fig-clusternet}. The network is unique in that different clusters automatically identify regions in the data that behave according to different functions (the function networks), a concept that is similar to that of local reduced-order bases, which are known to greatly improve model order reduction techniques for highly nonlinear dynamical systems~\cite{LROM}. A separate part of the architecture determines when and by how much to turn the other parts on or off (the context networks). Different loss functions are used for each part of the network to train them as either function or context networks. The separation in intent between different networks is similar to a mixture of experts network \cite{shazeer2017outrageously} and similar in construction to an AI Physicist network \cite{wu2018toward}. Because each cluster applies to a different function, and because each type of network has a separate role, the network is designed to be easier to interpret than a fully connected network. This network design makes an additional inductive bias over the smooth functions assumed by a fully connected network---that solutions can be broken down into a few regions, defined locally by very simple, smooth functions. 

This network requires a specialized training procedure with the function networks trained by a classification loss function, and the context networks trained successively by a regression loss function. Using the labels in Figure~\ref{fig-clusternet}, the training procedure for a network with two clusters is (1) train the function networks using a classification loss function, defined as the minimum of the absolute values $abs(f-f_1)$ and $abs(f-f_2)$ (2) train the cluster networks while holding the function network weights constant using a regression loss function, defined as the absolute value $abs(f-y)$.

\subsection{Fully Connected Network} 

\begin{figure}[hbt!]
\begin{center}
\centerline{\includegraphics[width=0.32\columnwidth]{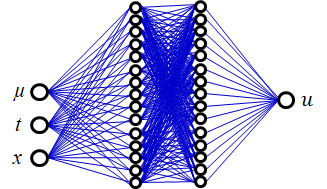}}
\caption{Fully connected network.}
\label{fig-fcn}
\end{center}
\vskip -0.2in
\end{figure}

The second proposed architecture, for comparison, is the simple feed-forward network in Figure~\ref{fig-fcn}. The network performs a mapping directly from a batch of ($x, t, \mu$) values to a batch of $u$ values where $x$ represents space, $t$ represents time, and $u$ represents the output of the network. For the Burgers' equation test case considered in this work, $u$ represents the velocity. For the shock-induced bubble test case also considered here, $u$ may represent the density, x-momentum, y-momentum, or energy. Also, $\mu$ can represents any design variable or parameter in the governing equation than can be easily modified from one simulation to the next. For the Burgers' equation test case, $\mu$ represents viscosity. For the shock-induced bubble test case, the free-stream Mach number $M$ is varied instead of $\mu$. Because $x$ and $t$ can be any scalar value, like the cluster network, this method is easy to apply to any time stepping procedure, grid type, or mesh resolution. 

\section{Datasets}

Because industry CFD simulations are expensive to compute, our datasets are designed to evaluate the network on its ability to learn well from very few examples. A dataset consisting of 40 simulations, but with variations in 5-10 design parameters would be a typical industry dataset. We test each method on a slightly harder version of the problem, scaled down in order to facilitate faster network evaluation. The networks are given the challenge of learning from only three example simulations with variation in a single design parameter for the two datasets. These datasets we have chosen to constrain and evaluate are the following Burgers' equation and shock-induced bubble test cases.

\subsection{Burgers' Equation Test Case}

\begin{figure}[hbt!]
\begin{center}
\centerline{\includegraphics[width=0.5\columnwidth]{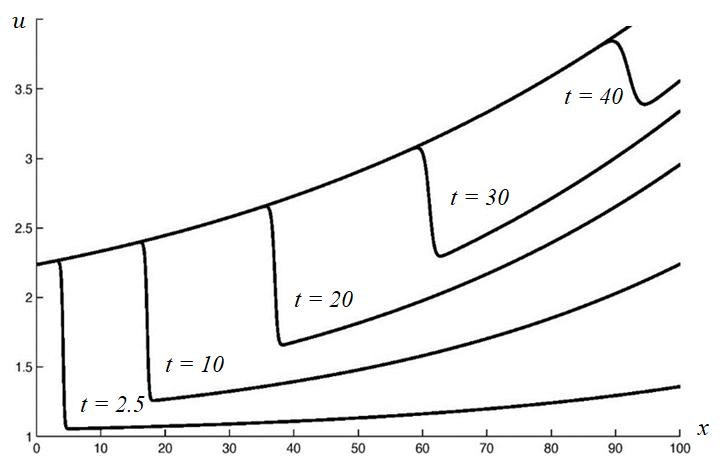}}
\caption{Burgers' equation evolution in time from 2.5 seconds to 40 seconds showing one full solution in time for a given set of parameters.}
\label{fig-burgers}
\end{center}
\vskip -0.2in
\end{figure}

The Burgers' equation test case shown in Figure~\ref{fig-burgers} is commonly used for evaluating ROMs. It is a one-dimensional application of an initial-boundary-value problem, which models the movement of a shockwave across a tube. 

\subsection{Shock-Induced Bubble Test Case}

The shock-induced bubble test case is a 2D application of the compressible Navier-Stokes equations, modeling a shockwave moving across a circular high density region representing a 2D bubble. In order to generate the dataset, parameters in the governing equations were varied within a CFD solver, including the free-stream Mach number ($M$), ratio of specific heats, and viscosity. Solutions were generated resolving 0.2 seconds in real time. 

\begin{figure}[hbt!]
\begin{center}
\centerline{\includegraphics[width=0.5\columnwidth]{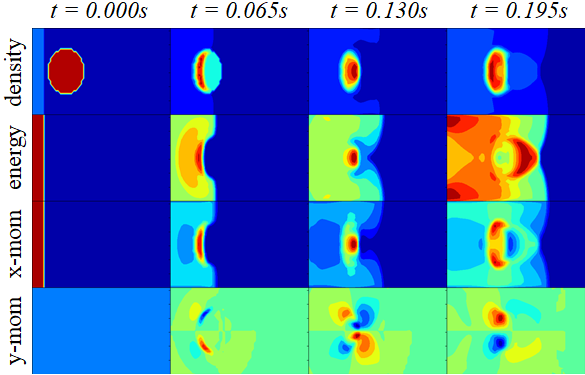}}
\caption{Shock-induced bubble contour plots showing the progression of the solution for its four state variables---density, energy, x-momentum, and y-momentum---at four points in time. Each solution resolves the progression of the shockwave over the bubble between t = 0.000s and 0.200s in 201 time steps with step size 0.001s. }
\label{fig-bubble-t200}
\end{center}
\vskip -0.2in
\end{figure}

At the beginning of the simulation---0.000 seconds---the problem is defined by a high density ``bubble'' with 5 times the density of the ambient fluid, as visualized in the left-most frames of Figure~\ref{fig-bubble-t200}. A moving shockwave is initialized near the left side of the simulation, represented by a small region with high x-momentum, which will move to the right and cross over the bubble. A Mach number greater than 1.0 indicates that the shockwave itself moves at higher than supersonic speed. The four state variables---density, x-momentum, y-momentum, and energy are plotted at t = 0.000s through t = 0.195s in Figure~\ref{fig-bubble-t200} for a shockwave Mach number of 1.8. 

\section{Numerical Results}

\subsection{Burgers' Equation Results}

\subsubsection{Results Compared to Baselines}

Results for all baseline and proposed methods are tabulated in Table~\ref{method-rmse-table} for the Burgers' equation test case. ROMs, fully connected networks and cluster networks all predict solutions at unseen conditions well, and predict the QoIs calculated from the full solutions well. The fully connected network and cluster network also obtain predictions about two orders of magnitude faster than the full order model. 

\begin{table}[hbt!]
\caption{Burger's equation test case: performance comparisons with baselines. Online CPU time, percent error on the test set, and percent error on selected quantities of interest are shown for different methods compared to the full order model. Abbreviations: Full order model (FOM), ROM with 100 basis vectors (ROM 100), Fully Connected Network with 4 layers and 25 nodes per layer (FCN 4 20), Cluster Network with 2 clusters 3 layers 5 nodes per layer (ClusterNet 2 3 5), Surface fitting using 1st order polynomial on a QoI (QoI Poly Fit 1). }\label{method-rmse-table}
\begin{center}
\begin{small}
\begin{tabular}{lrrrrr}  
\hline
Method & CPU (s)  & \% Error & QoI \% Error \\
\hline
FOM              &      16.273&       &        \\
ROM 100       &      12.981& 0.07& 0.05 \\
ROM 50         &       5.690 & 0.20& 0.16\\
ROM 20         &       3.575 & 0.86& 0.60\\
ROM 5           &       2.415 & 4.04& 3.00\\
ROM 3           &       2.119 & 5.47& 5.28\\
FCN 4 20        &      0.245 & 2.43& 3.04\\
ClusterNet 2 3 5&      0.227 & 0.71& 0.42\\
QoI Poly Fit 3 & $<$0.001 &       & 7.28 \\
QoI Poly Fit 2 & $<$0.001 &       & 7.15 \\
QoI Poly Fit 1 & $<$0.001 &       & 8.42 \\
QoI Poly Fit 0 & $<$0.001 &       &15.03 \\
\hline
\end{tabular}
\end{small}
\end{center}
\end{table}

Some QoIs were constructed for more meaningful comparisons because in practice there are usually QoIs that can be computed from a simulation that are important to a design engineer in addition to the full solution. Therefore, four psuedo-QoI's representative of the kinds of quantities engineers might investigate were calculated from the test set (1) a single random solution point in space and time (2) the average value of the solution at a random fixed point in time (3) the average value of the solution at a random fixed point in space and (4) the average value of the solution over space and time. Similar QoIs were constructed for both Burgers' equation and the shock-induced bubble test case. The final tabulated QoI is the average of the four psuedo-QoIs.

The percent error is calculated over the entire computational domain by (1) scaling the values for all solutions in the dataset to between 0 and 1, (2) taking the absolute value of the difference between the predicted solutions and solutions computed by the full order model, and (3) multiplying by 100. This percent error metric provides a measure of the difference between the prediction and the ground truth, represented as a percent of the maximum value after the variables have been scaled between 0 and 1. The percent errors are the same as the scaled L1 errors typically investigated in machine learning applications, except multiplied by 100. The QoI percent error is calculated with the same procedure, except in step (2) we take the absolute value of the difference between the predicted QoI and the QoI computed by the full order model. 

\subsubsection{Visualization of Fully Connected Network Results}

\begin{figure}[hbt!]
\begin{center}
\centerline{\includegraphics[width=0.42\columnwidth]{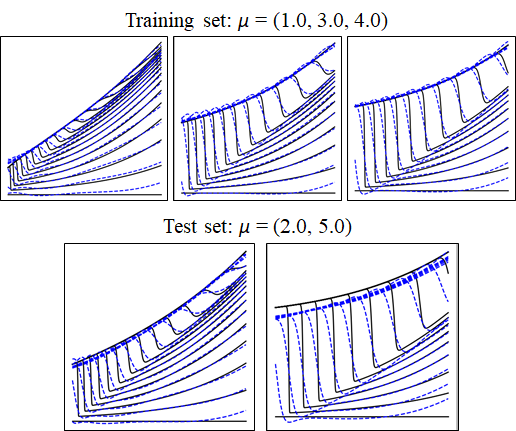}}
\caption{Fully connected network results on Burgers' equation.}
\label{fig-burg-fcn-results}
\end{center}
\vskip -0.4in
\end{figure}

With the right hyper-parameter tuning and the implementation of exponential learning rate decay, the single fully connected network learns to fit the data and predict solution behavior at unseen parameters as shown in Figure~\ref{fig-burg-fcn-results}. Note that the method not only interpolates, but extrapolates to some extent beyond parameters it has seen---this is a surprising result as we only expected neural networks to interpolate well, and the method is only required to interpolate well to be part of a design optimization procedure. 

\subsubsection{Visualization of Cluster Network Results}

\begin{figure}[hbt!]
\begin{center}
\centerline{\includegraphics[width=0.42\columnwidth]{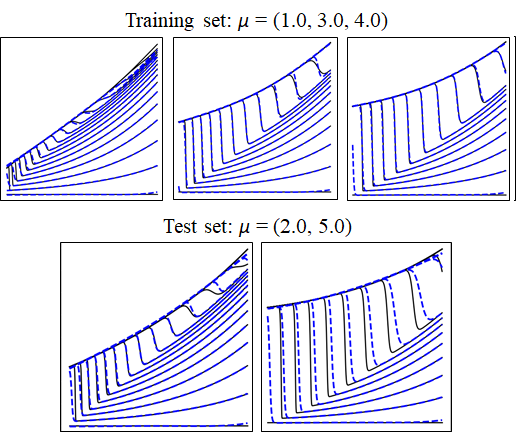}}
\caption{Cluster network results on Burgers' equation.}
\label{fig-burg-clusternet-results}
\end{center}
\vskip -0.4in
\end{figure}

The cluster network model was also trained on only three examples, and tested on two examples. Overlays of CFD solutions with neural network predictions in Figure~\ref{fig-burg-clusternet-results} show that the neural network predictions closely approximate the full order model solutions, interpolating and extrapolating better than the fully connected network. 

\subsection{Shock-Induced Bubble Results}

\subsubsection{Results Compared to Baselines}

We compare our shock-induced bubble results to the full order model and the QoI baselines in Table~\ref{method-bubble-table}. Again, the fully connected network and the cluster network outperform the model-free surface fitting methods. The neural network-based approach has an even greater speedup over the full order model on higher-dimensional problems. 

\begin{table}[hbt!]
\caption{Shock-induced bubble test case: performance comparisons with baselines. Online CPU time and percent error on selected QoIs from the test set shown for different methods compared to the full order model. Abbreviations: Full order model (FOM), Fully Connected Network with 4 layers and 40 nodes per layer (FCN 4 40), Cluster Network with 4 clusters 3 layers 15 nodes per layer (ClusterNet 4 3 15), Surface fitting using 2nd order polynomial fit on a quantity of interest (QoI Poly Fit 2). }
\begin{center}
\begin{small}
\begin{tabular}{lrrrrr}
\hline
Method & CPU (s)  & QoI \% Error \\
\hline
FOM                  & 98.374      &        \\
FCN 4 40            & 0.687       & 0.37 \\
ClusterNet 4 3 15  & 0.524       & 0.88 \\
QoI Poly Fit 3     & $<$0.001 &  5.73 \\
QoI Poly Fit 2     & $<$0.001 &  5.29 \\
QoI Poly Fit 1     & $<$0.001 &  1.78 \\
QoI Poly Fit 0     & $<$0.001 &  1.79 \\
\hline
\end{tabular}
\end{small}
\end{center}
\vskip -0.3in
\label{method-bubble-table}
\end{table}

\subsubsection{Visualization of Results}

\begin{figure}[hbt!]
\begin{center}
\centerline{\includegraphics[width=0.75\columnwidth]{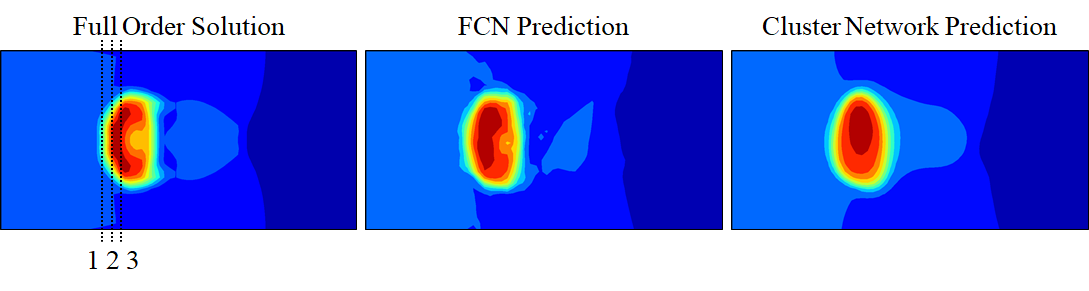}}
\caption{2D density contours for the fully connected network and cluster network compared to the full order solution for the shock-induced bubble for one test set example at t = 0.18s and M = 1.8.}
\label{fig-bubble-example-results}
\end{center}
\vskip -0.2in
\end{figure}

\begin{figure}[hbt!]
\begin{center}
\centerline{\includegraphics[width=0.62\columnwidth]{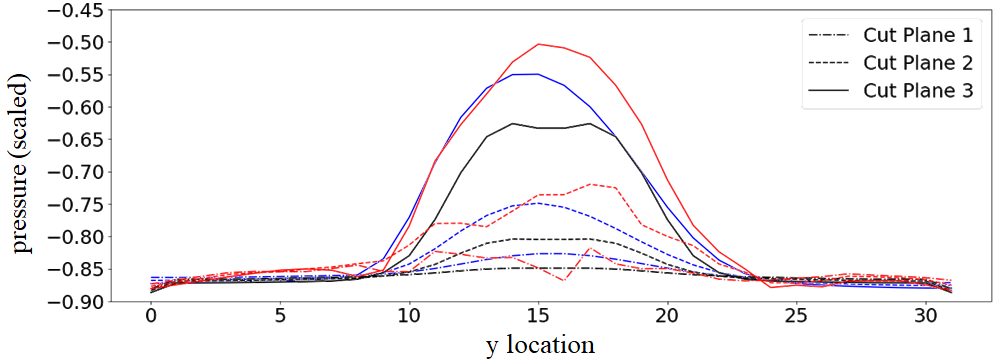}}
\caption{Density contours for the fully connected network predictions (red), cluster network predictions (blue), and the full order solutions (black) for the shock-induced bubble for one example in the test set at t = 0.18s and M = 1.8 along three cut planes taken near the shockwave at the dashed lines in Figure~\ref{fig-bubble-example-results}.}
\label{fig-bubble-contour-results}
\end{center}
\vskip -0.2in
\end{figure}

We visualize the accuracy of the network for the shock-induced bubble at a time point near the end of the simulation at 0.180 seconds. The density is shown on a contour plot in $x$ and $y$ for each of the shockwave Mach numbers in the training and test sets for the full order CFD solutions and the neural network predictions. 
The contour plots in Figure~\ref{fig-bubble-example-results} show that this relatively small network predicts the solutions at free-stream Mach numbers that it never saw with some signicant errors at the locations with more strongly nonlinear density profiles. However, Table~\ref{method-bubble-table} reveals that its overall accuracy remains competitive with other model-free methods. Figure~\ref{fig-bubble-contour-results} shows the shape of the density contours near the shockwave along the three dashed cut planes in Figure~\ref{fig-bubble-example-results}, where the prediction errors are the highest.  

\section{Conclusions}

Both neural network architectures tested are more accurate than conventional surface fitting. For the one-dimensional Burgers' 
equation, they are found to deliver the same accuracy as ROMs, but are an order of magnitude faster. Between the two networks, the fully connected network is a good baseline to use for comparison with other neural network architectures for a few reasons: (1) the architecture is simple, (2) the training procedure is simple and (3) the inductive bias assumes smooth functions. However, in practice the network requires extensive hyper-parameter search and tuning. The cluster network is more accurate than the fully connected network for the Burgers' equation test case, and it provides additional benefits: (1) it uses fewer weights, (2) the forward pass is faster than the fully connected network and (3) its inductive bias is more applicable to highly nonlinear CFD problems---a smooth local bias. Ultimately, both network architectures perform well, and the percent error is under 1\% on the two test cases for the most successful model---the cluster network.

\section{Future Work}

Future neural network architectures with inductive biases even better suited to fluid dynamics are likely to achieve more accurate results, and compete with model-based methods like ROMs across a wider range of problems. Additionally, demonstrations on high-dimensional industry test cases are necessary for our method to be adopted within industry for design and optimization. Finally, the solution-predicting is only part of the optimization procedure. In order for these tools to ultimately be useful for design and optimization, they will need to be paired with a process for mapping out how quantities of interest vary over the design parameter space. Future work would require a method for choosing the next points for running the expensive simulations, similar to the ``greedy'' selection procedure used in ROMs, which balances exploration and exploitation. With the addition of a point selection method, we would have a complete procedure for engineers to use for aerodynamic design optimization. 

\section{Acknowledgments}

This first author acknowledges the support by the Department of Energy Computational Science Graduate Fellowship (DOE CSGF) program through US DOE Grant DE-FG02-97ER25308. 
The third author acknowledges partial support by the Air Force Office of Scientific Research under grant FA9550-17-1-0182, and partial support by a research grant from the King
Abdulaziz City for Science and Technology (KACST). At Lawrence Berkeley National Laboratory, Ann Almgren provided the AMReX full order model code used to generate the shock-induced bubble test case and Matt Zahr provided the pyMORTestbed full order model and reduced order model code used to generate and evaluate the Burgers' equation test case. We are grateful to Ann Almgren and Nathaniel Thomas for text revisions and insights. 

\bibliography{sample}

\end{document}